\documentclass[a4paper,11pt]{article}
\usepackage{jinstpub}

\usepackage{physics}
\usepackage{mathtools}
\usepackage{braket}
\usepackage{textgreek}
\usepackage{xcolor}
\usepackage{siunitx}
\usepackage{comment}
\usepackage{multirow}
\usepackage{array}
\usepackage{booktabs}
\usepackage{enumitem}

\usepackage{subcaption}

\title{\boldmath Pulse Shape Model for a Light Enhanced Phonon Detector}

\collaboration{The COSINUS Collaboration}
\author[a]{G.~Angloher,}
\author[a]{M.~R.~Bharadwaj,}
\author[b,c]{A.~B\"ohmer,}

\author[b,c,1]{S.~Braun,\note{Corresponding author.}}

\author[b,c]{M.~R.~Cababie,}
\author[d]{I.~Colantoni,}
\author[d,e]{I.~Dafinei,}
\author[e,f]{N.~Di~Marco,}
\author[a]{C.~Dittmar,}
\author[f]{F.~Ferella,}
\author[d,e]{F.~Ferroni,}
\author[b]{S.~Fichtinger,}
\author[g,f]{A.~Filipponi,}
\author[b]{M.~Friedl,}
\author[b,c]{D.~Fuchs,}
\author[h]{L.~Gai,}
\author[a]{M.~Gapp,}
\author[i]{M.~Heikinheimo,}
\author[a,m]{K.~Heim,}
\author[a]{M.~N.~Hughes,}
\author[i]{K.~Huitu,}
\author[b,c]{M.~Kellermann,}
\author[a]{M.~Mancuso,}
\author[e,f]{L.~Pagnanini,}
\author[a]{F.~Petricca,}
\author[e,f]{C.~Petrucci,}
\author[f]{S.~Pirro,}
\author[a]{F.~Pr\"obst,}
\author[g,f]{G.~Profeta,}
\author[f]{A.~Puiu,}
\author[b,c]{F.~Reindl,}
\author[a,m]{K.~Sch\"affner,}
\author[b,c]{J.~Schieck,}
\author[b,c]{P.~Schreiner,}
\author[b,c]{C.~Schwertner,}
\author[j]{P.~Settembri,}
\author[a]{K.~Shera,}
\author[a]{M.~Stahlberg,}
\author[i]{A.~Stendahl,}
\author[k]{M.~Stukel,}
\author[f,l]{C.~Tresca,}
\author[h]{S.~Yue,}

\author[a,b,2]{V.~Zema,\note{Corresponding author.}}

\author[h]{Y.~Zhu,}
\author[a]{L.~Ziegele,}
\author[i]{N.~Zimmermann,}
\author[a]{\\ and O.~Schulz}

\affiliation[a]{Max-Planck-Institut f\"ur Physik,\\
85748 Garching, Germany}

\affiliation[b]{Marietta-Blau-Institut f\"ur Teilchenphysik,\\
1010 Wien, Austria}

\affiliation[c]{Atominstitut, Technische Universit\"at Wien,\\
1020 Wien, Austria}

\affiliation[d]{INFN -- Sezione di Roma,\\
00185 Roma, Italy}

\affiliation[e]{Gran Sasso Science Institute,\\
67100 L'Aquila, Italy}

\affiliation[f]{INFN -- Laboratori Nazionali del Gran Sasso,\\
67100 L'Aquila, Italy}

\affiliation[g]{Dipartimento di Scienze Fisiche e Chimiche,
Universit\`a degli Studi dell'Aquila,\\
67100 L'Aquila, Italy}

\affiliation[h]{SICCAS -- Shanghai Institute of Ceramics,\\
Shanghai 201899, P.R. China}

\affiliation[i]{Helsinki Institute of Physics, University of Helsinki,\\
00014 Helsinki, Finland}

\affiliation[j]{PSI Center for Scientific Computing, Theory and Data,
Paul Scherrer Institute,\\
5232 Villigen PSI, Switzerland}

\affiliation[k]{SNOLAB,\\
Lively, Ontario P3Y~1N2, Canada}

\affiliation[l]{CNR-SPIN, c/o Dipartimento di Scienze Fisiche e Chimiche,
Universit\`a degli Studi dell'Aquila,\\
67100 L'Aquila, Italy}
\affiliation[m]{Physik-Department, TUM School of Natural Sciences, Technische Universität München, \\85747 Garching, Germany}

\emailAdd{sarah.braun@oeaw.ac.at}
\emailAdd{vanessa.zema@mpp.mpg.de}
\abstract{
COSINUS has tested prototype NaI-based cryogenic scintillating calorimeters with a composite design, in which the transition edge sensor (TES) is fabricated on a separate carrier crystal coupled to the NaI target. Measurements with the composite detectors showed deviations {\color{black}from the known pulse shape description for cryogenic calorimeters} and did not reveal an identifiable nuclear recoil population {\color{black}during a} neutron calibration. In this work we present a new pulse shape model which provides an explanation to these observations and that laid the foundation for the development of the final COSINUS phonon detector design named remote TES (remoTES). We demonstrate that the first COSINUS detector prototypes featured a phonon channel with {\color{black}signal enhancement} by light absorption in the carrier, namely a \textit{light enhanced phonon detector}. {\color{black} This design offers for the first time the chance to use a single channel phonon detector for particle discrimination.} We present recent results using a novel pulse shape fitting method based on the Bayesian inference software package BAT.jl.
}

\begin{document}
\maketitle
\flushbottom

\section{Introduction}
\begin{figure}[ht]
\begin{centering}
\includegraphics[width=0.8\linewidth]{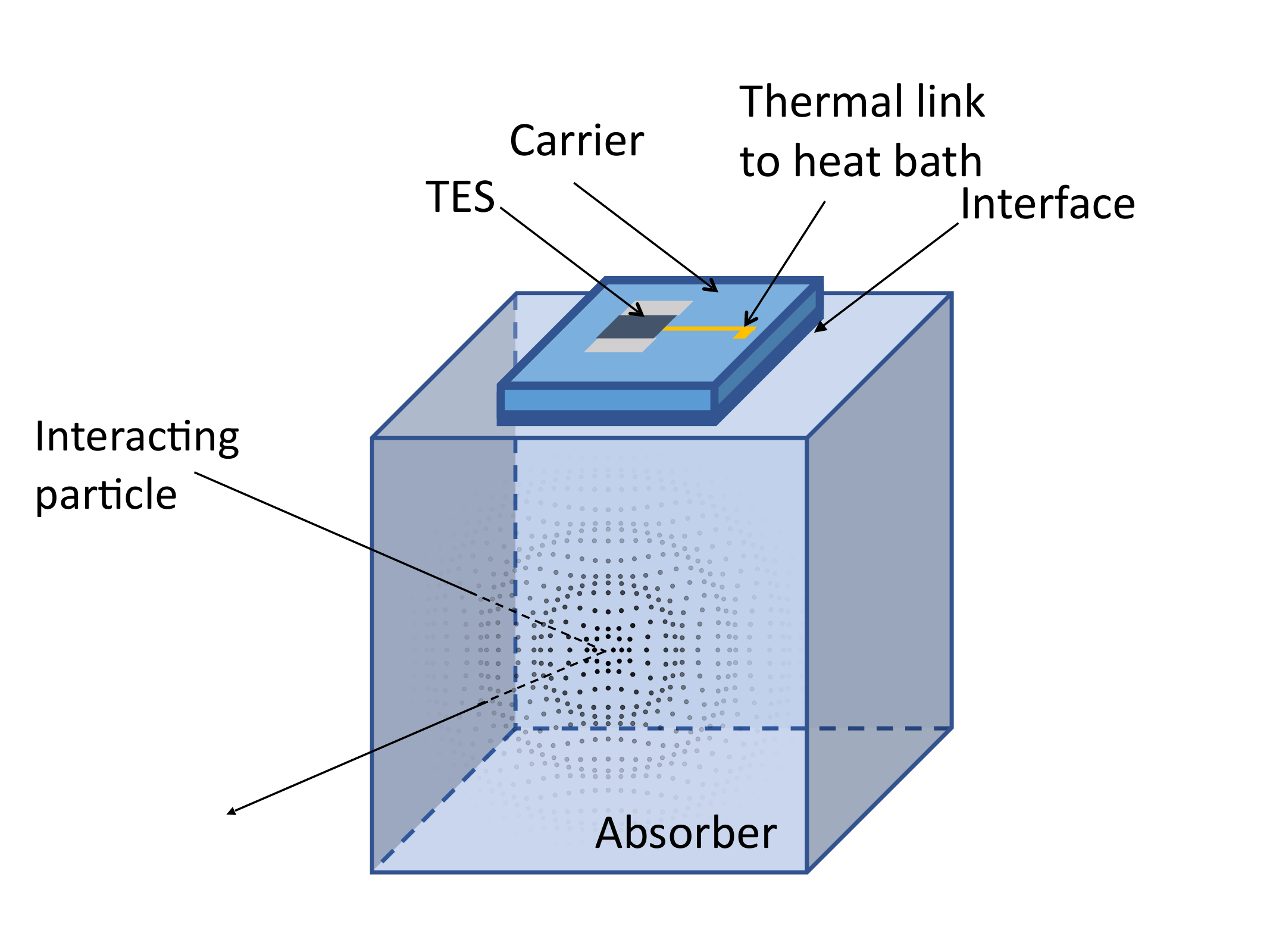}
\caption{\small Scheme of a composite design for a cryogenic calorimeter.}
\label{fig:composite_scheme}
\end{centering}
\end{figure}

Cryogenic calorimeters are detectors widely employed for rare event searches in particle physics, e.g. for dark matter and neutrino scattering detection, because of their exceptional energy resolution at low energies~\cite{Kim:2022lqb, COSINUS:2023kqd, CRESST:2024cpr, CUORE:2022aiq, Trotta:2024cfl, Lattaud:2022jnq, NUCLEUS:2019igx, Ricochet:2023nvt, SuperCDMS:2020aus, Kennard:2026xlx, TESSERACT:2025tfw}. {\color{black}They consist of a selected target material, the \textit{absorber}, coupled to} an ultra sensitive temperature device fabricated on its surface, the \textit{sensor}. They are employed to measure the energy deposited by a particle interaction in the absorber and converted into phonons. About 30 years of development of TES-based cryogenic calorimeters have been carried out at the Max Planck Institute for Physics, starting from the work of W. Seidel et al.~\cite{SEIDEL1990483} which led to the foundation of the CRESST~\cite{CRESST:2024cpr, CRESST:2022dtl, CRESST:2019jnq}, COSINUS~\cite{Angloher:2025shf} and NUCLEUS~\cite{NUCLEUS:2019igx} experiments. The refined fabrication process of tungsten TES (W-TES) produces {\color{black}detectors that are among the most sensitive  cryogenic calorimeters} for rare event searches worldwide~\cite{CRESST:2024cpr, CRESST:2022dtl, CRESST:2019jnq}. Some target materials are considered unsuitable for cryogenic applications because of their softness, low melting point or hygroscopicity. For instance, NaI must be stored and handled only in {\color{black} air-tight containers with controlled nitrogen atmosphere} and it cannot withstand the evaporation of a W-TES because of the high temperature required~\cite{Angloher:2017sft}. Nevertheless, the COSINUS experiment is developing NaI cryogenic \textit{scintillating} calorimeters (NaI-CSCs) to cross-check the results of the DAMA/LIBRA experiment using the same target material and an event-by-event discrimination of {\color{black}nuclear recoils} from other events~\cite{Angloher:2017sft}. To this aim, in an early stage, COSINUS adopted a workaround and evaporated the TES on a different crystal, named \textit{carrier}. The carrier was placed in thermal contact with the absorber, e.g. by gluing it on the absorber surface~\cite{Angloher:2017sft}, see Fig.~\ref{fig:composite_scheme}. This \textit{composite design}, which also preserves the absorber from the TES fabrication process, was previously successfully employed by CRESST-II~\cite{CRESST:2015txj}, but using the same material for absorber and carrier, i.e. CaWO$_4$. 

The combination of an absorber made from NaI and a carrier of a different material (e.g. CaWO$_4$, CdWO$_4$, {\color{black} Al$_2$O$_3$}, Si) led to two observations in the COSINUS composite prototypes: a different pulse shape with respect to the reference model for this type of detectors (Pr\"obst et al.~\cite{probst1995model}, see appendix) and the lack of {\color{black} identification of nuclear recoil events in the} data taken during irradiation with neutron sources. {\color{black} Since nuclear recoils produce less scintillation light than electron recoils due to quenching, they are expected to populate a separate band in the light yield versus energy plane.} {\color{black}As discussed in this work, both observations can be understood as a consequence of scintillation light absorption in the carrier crystal. This insight motivated the transition from the carrier-based composite prototype concept to the final COSINUS detector design, the remoTES \cite{easy_to_fabricate}.}

The focus of the present work, however, is the interpretation of the earlier carrier-based composite prototypes. We show that these detectors can be understood as \textit{light-enhanced phonon detectors}, in which the TES measures a superposition of the absorber-phonon signal and an additional phonon signal generated by scintillation-light absorption in the carrier. In this way, an accidental experimental finding is turned into an opportunity for single-channel particle discrimination. To describe the observations made with COSINUS composite prototypes, we extend the thermal model of Pröbst et al.~\cite{probst1995model}, originally developed for cryogenic calorimeters. In Sec.~\ref{sec:detectors}, we detail the design of two composite prototypes, NaIce2 and NaIce6. In Sec.~\ref{sec:observation}, we describe the experimental observations reported by COSINUS which motivated the extension of the model explained in Sec.~\ref{sec:modeldesc}. In Sec.~\ref{sec:LEPD}{\color{black}, we present a pulse shape model fit to the NaIce2 and NaIce6 data based on Bayesian inference, which validates the model, demonstrates its effectiveness for neutron-event discrimination, and yields new insights on the detector response. We conclude in Sec.~\ref{sec:conclusions}.}

\section{Detector description}
\label{sec:detectors}
\begin{figure}[htbp]
   \centering
   \includegraphics[width=\linewidth]{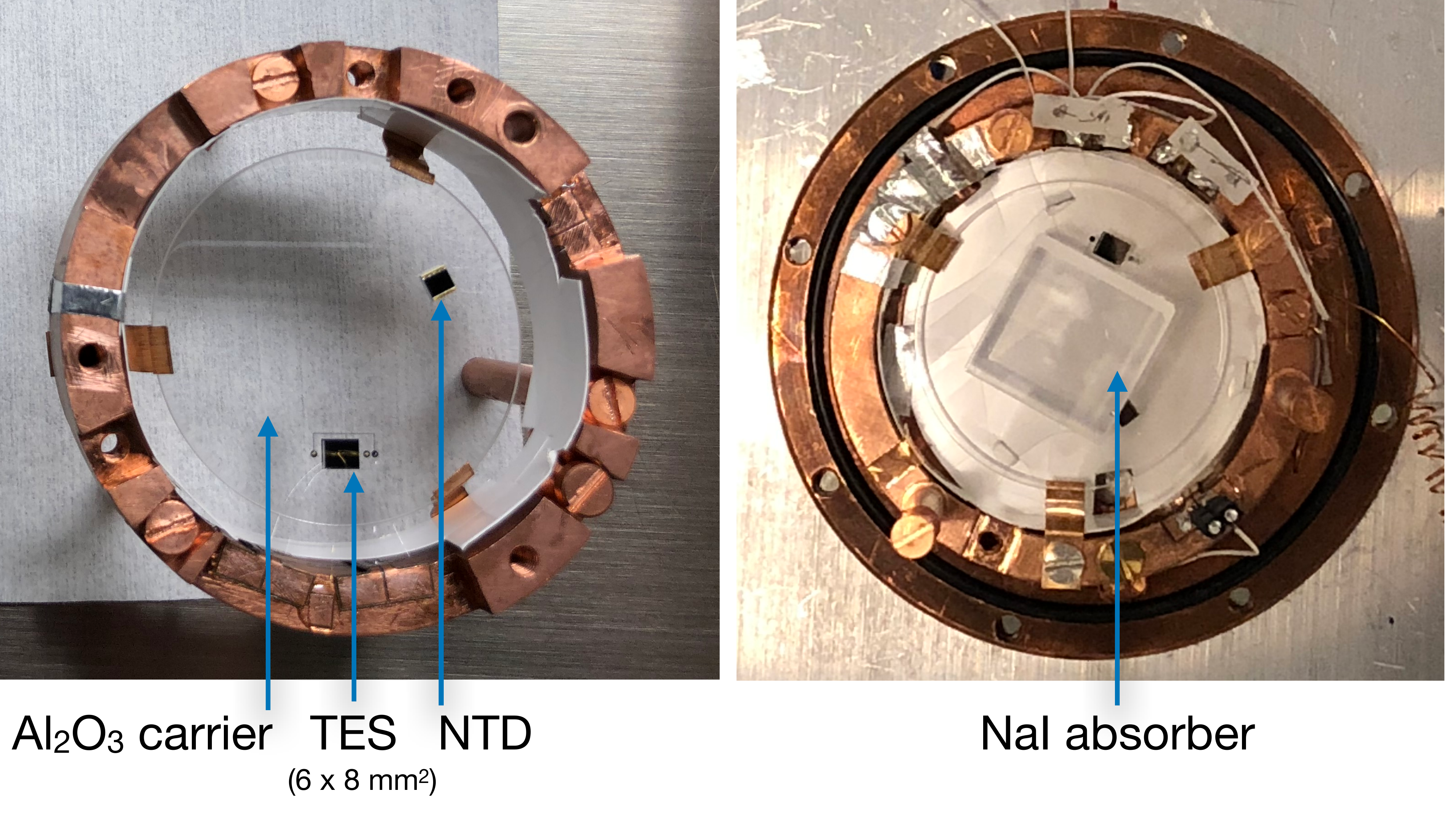}
   \caption{Picture of the NaIce6 composite design detector. Left: Al$_2$O$_3$ carrier, equipped with both a TES and an NTD sensor and mounted into a copper holder. Right: NaI absorber glued to the sapphire crystal.}
   \label{fig:NaIce6}
\end{figure}

{\color{black} In this work we focus on two composite design detectors equivalent to the one described in~\cite{Angloher:2017sft}, dubbed NaIce2 and NaIce6~\cite{Zema:2020mkm}. 
NaIce2 is a $(30\times30\times20)$ mm$^3$ pure NaI crystal, produced by the Hilger company and glued to a CdWO$_4$ crystal carrier of 39.2 mm diameter and 1.6 mm thickness. The temperature sensor is a $(6 \times 8)$ mm$^2$ W-TES, 200 nm thick. {\color{black} It consists only of W, and does not feature the two aluminum (Al) films which are typically used as phonon collectors by, e.g., CRESST}. The thermal link to the copper holder is obtained using a Au-wire of \SI{25}{\micro\metre}  diameter. 
NaIce6 is a $(30\times30\times20)$ mm$^3$ pure NaI crystal, produced by the Hilger company and glued to an Al$_2$O$_3$ crystal carrier of 40 mm diameter and 0.45 mm thickness. The W-TES and the thermal link are equivalent to the ones of NaIce2. Additionally, a Neutron Transmutation Doped (NTD) sensor is fabricated on the Al$_2$O$_3$ crystal carrier. The description of the NaIce2 and NaIce6 modules is summarized in Tab.~\ref{tab:detector_design}. Pictures of NaIce6 are shown in Fig.~\ref{fig:NaIce6}.}

\begin{table}[ht]
\caption{ Description of the design of the composite detectors considered in this work. $D$ denotes diameter, $t$ thickness, and TL the thermal link.}

\centering

\begin{tabular}{l c c c c c}
\hline
 & Absorber & Interface & Carrier & Sensor & TL \\
\hline

NaIce2 &
\begin{tabular}{@{}c@{}}NaI\\$(30\times30)$ mm$^2$\\$t$: \SI{20}{\milli\metre}\end{tabular} &
EpoTek 301-2 &
\begin{tabular}{@{}c@{}}CdWO$_4$\\$D$: \SI{39.2}{\milli\metre}\\$t$: \SI{1.6}{\milli\metre}\end{tabular} &
\begin{tabular}{@{}c@{}}W-TES\\$(6\times8)$ mm$^2$\\$t$: \SI{200}{\nano\metre}\end{tabular} &
\begin{tabular}{@{}c@{}}Au-wire\\$D$: \SI{25}{\micro\metre}\end{tabular} \\

\hline

NaIce6 &
\begin{tabular}{@{}c@{}}NaI\\$(20\times20)$ mm$^2$\\$t$: \SI{20}{\milli\metre}\end{tabular} &
EpoTek 301-2 &
\begin{tabular}{@{}c@{}}Al$_2$O$_3$\\$D$: \SI{40}{\milli\metre}\\$t$: \SI{0.45}{\milli\metre}\end{tabular} &
\begin{tabular}{@{}c@{}}W-TES\\$(6\times8)$ mm$^2$\\$t$: \SI{200}{\nano\metre}\\+ NTD\end{tabular} &
\begin{tabular}{@{}c@{}}Au-wire\\$D$: \SI{25}{\micro\metre}\end{tabular} \\

\hline
\end{tabular}
\label{tab:detector_design}
\end{table}


\section{Experimental observations}

\label{sec:observation}
COSINUS reported on two observations:

\begin{enumerate}

\item \textit{The general pulse shape model in}~\cite{probst1995model}\textit{ did not fit the NaI composite detector pulse shape.} 
A measurement involving NaIce2 was performed in the cryogenic test facility of the Max Planck Institute for Physics located in the underground National Laboratories of Gran Sasso (LNGS) and the results are reported in~\cite{Angloher:2017sft}. An accurate analysis of the pulse shape of a standard event, obtained by averaging the pulses populating the \SI{60}{\kilo\electronvolt} X-ray peak of the $^{241}$Am calibration source, showed that the general model for this kind of detectors~\cite{probst1995model} did not fit the pulse shape. An additional empirical exponential term was incorporated into the general model of~\cite{probst1995model} to describe the pulse shape of NaIce2.

\item \textit{Neutron events could not be discriminated from $e^{-}/\gamma$ events.} 
CSCs data consist in the coincident readout of light and phonon signals produced by a primary interaction in the target. For the same deposited energy in the absorber, {\color{black} the interaction with electrons generates more scintillation light than the elastic scattering off nuclei}, according to the phenomenon of light quenching~\cite{Birks:1964zz}. We define the \textit{light yield ($LY$)} as the ratio between the energy detected in the light channel and the one in the phonon channel. The $LY$ vs total energy plot is expected to show $e^{-}/\gamma$ and neutron events separated and populating different bands. This procedure is validated by several CRESST measurements, see e.g. \cite{CRESST:2019jnq}, {\color{black} and recent results obtained with COSINUS prototypes with the NaI remoTES design}~\cite{COSINUS:2023qwi, COSINUS:2023kqd}. When the COSINUS composite detector was {\color{black} paired with} a light detector and the $LY$ calculation was applied to data taken during irradiation with a neutron source, the expected identification of the {\color{black} nuclear recoil} events was not possible, as no {\color{black} quenched bands were observed. To investigate this observation, NaIce6 was equipped with both a TES and an NTD sensor~\cite{McCammon2005} and surrounded by a silicon beaker-shaped light detector to discriminate $e^-/\gamma$ events from nuclear recoils. The analysis of data recorded with the NTD in coincidence with the light detector allowed to tag the neutrons.}
\end{enumerate}

\begin{figure}[htbp]
   \centering
   \includegraphics[width=0.6\linewidth]{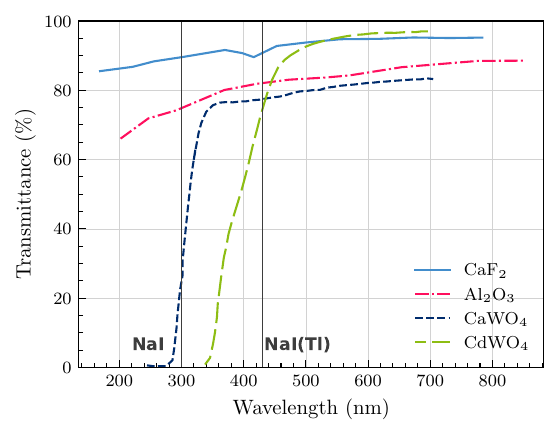}
   \caption{ Crystal light transmittance as a function of the wavelength for CaF$_2$ (blue)~\cite{driscoll1978handbook}, Al$_2$O$_3$ (magenta), CaWO$_4$ (dark blue)~\cite{ninkovic2005investigation} and CdWO$_4$ (green)~\cite{Danevich:2013xzw} . For pure NaI and NaI(Tl) in addition the peaks of the scintillation light emission are shown, see~\cite{creutz2012nuclear}.}
   \label{fig:transmission}
\end{figure}

{\color{black} The} materials used for the carriers are not transparent to the NaI scintillation light~\cite{Danevich:2013xzw,driscoll1978handbook,ninkovic2005investigation,creutz2012nuclear}. In Fig.~\ref{fig:transmission}, we report the nominal transmission probability curves of materials employed for the carriers as a function of wavelength~\citep{Danevich:2013xzw, driscoll1978handbook, ninkovic2005investigation}. We indicate the nominal wavelengths of NaI and NaI(Tl) scintillation light, corresponding to \SI{300}{\nano\metre} and \SI{430}{\nano\metre} \cite{creutz2012nuclear}, respectively. Assuming these values are valid for the two modules considered in this work, the transmission curves show that in NaIce2 and NaIce6 the carriers are expected to partially absorb the NaI scintillation light. Although the actual wavelength changes from crystal to crystal because it depends on many factors such as crystal structure, level of doping and temperature \cite{Rodnyi:1997,Mikhailik:2006}, the possible spread does not change the expectation that the carriers are opaque to the scintillation light of the NaI absorber.

These observations strongly suggested that the pulse shape of the composite detector in Fig.~\ref{fig:composite_scheme} is affected by the absorption of the NaI scintillation light in the carrier. Specifically, we considered the hypothesis that the pulse detected in the phonon channel is the superposition of a pulse produced by the direct particle hits in the absorber and the pulse produced by the scintillation light absorption in the carrier.

\begin{figure}
\begin{centering}
\includegraphics[width=0.5\linewidth]{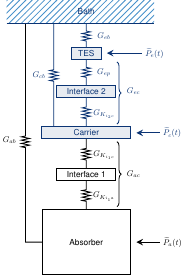}
\caption{
Thermal-coupling scheme of the composite cryogenic calorimeter,
consisting of an absorber, carrier, and TES. The blue-highlighted
subsystem corresponds to the original two-component model of
Pröbst et al.~\cite{probst1995model}: in the limit in which the absorber and its couplings are removed, the carrier takes the
role of the absorber in the original model. The $G_{jk}$ denote thermal
conductances between components $j$ and $k$; $G_{\mathrm{K}_{ij}}$
denote Kapitza (thermal-boundary) conductances between interface $i$
and component $j$; and $G_{\mathrm{ep}}$ denotes the electron-phonon coupling. $\tilde{P}_{\mathrm{a}}(t)$,
$\tilde{P}_{\mathrm{c}}(t)$, and $\tilde{P}_{\mathrm{e}}(t)$ are the
power inputs to the absorber, carrier, and TES, respectively.}
\label{fig:det_scheme_carrier}
\end{centering}
\end{figure}

\section{Pulse shape model} \label{sec:modeldesc}

{\color{black}The model distinguishes between non-thermal and thermal phonon populations. We refer to \emph{non-thermal} phonons as the prompt high-frequency phonon population produced after an energy deposition by particles. These phonons are not yet in thermal equilibrium with the crystal lattice and undergo successive anharmonic decay and scattering while propagating through the detector. After sufficient down-conversion and multiple scattering events, the phonon distribution approaches thermal equilibrium. We refer to this equilibrated population as \emph{thermal} phonons, characterized by the temperature of the corresponding detector component.}

In Eq.~\eqref{eq:extendedsystem}, we present the system of differential equations which extends the model in~\cite{probst1995model} to include the coupling of the absorber and the sensor to the carrier and add the power input injected in the carrier by the scintillation light absorption. The thermal-coupling scheme of the
composite cryogenic calorimeter is shown in
Fig.~\ref{fig:det_scheme_carrier}.

\begin{equation}
\small
\label{eq:extendedsystem}
    \begin{cases}
\tilde{P}_{\mathrm{e}}(t) = &C_{\mathrm{e}} \frac{dT_{\mathrm{e}}}{dt} +  G_{\mathrm{ec}}\Delta T_{\mathrm{ec}} + G_{\mathrm{eb}}\Delta T_{\mathrm{eb}} \\
	\tilde{P}_{\mathrm{c}}(t)= & C_{\mathrm{c}} \frac{dT_{\mathrm{c}}}{dt} + G_{\mathrm{ec}}\Delta T_{\mathrm{ce}} + G_{\mathrm{ac}} \Delta T_{\mathrm{ca}} + G_{\mathrm{cb}} \Delta T_{\mathrm{cb}}  \\
	 \tilde{P}_{\mathrm{a}}(t)= &C_{\mathrm{a}} \frac{dT_{\mathrm{a}}}{dt} + G_{\mathrm{ac}}\Delta T_{\mathrm{ac}} + G_{\mathrm{ab}}\Delta T_{\mathrm{ab}} 
    \end{cases},
\end{equation}

where $C_j$ and $T_j$, for $j=\mathrm{e},\mathrm{c},\mathrm{a}$ are the TES (e), carrier (c) and absorber (a) heat capacities and temperatures, respectively, and {\color{black}$G_{jk}$ and $\Delta T_{jk}= T_j -T_k$ are the pairwise conductances and temperature differences between the components and between the components with the bath (b).} The power inputs $\tilde{P}_{j}$ are,
\begin{equation}
    \label{eq:powerinputs}
    \begin{cases}
	 \tilde{P}_{\mathrm{e}}(t) & =  P_0\; e^{-t /\tau'_{\mathrm{n}}} + P_\ell \; e^{-t/\tau_\ell}\\
	 \tilde{P}_{\mathrm{c}}(t) & = \frac{1-\varepsilon}{\varepsilon} \; P_0 \; e^{-t/\tau'_{\mathrm{n}}} + \frac{1-\xi}{\xi} \; P_\ell \; e^{-t/\tau_\ell}\\
	 \tilde{P}_{\mathrm{a}}(t) & =  \frac{1-\eta}{\varepsilon\;  \eta} \; P_0 \; e^{-t/ \tau'_{\mathrm{n}}}
    \end{cases},
\end{equation}

where $P_0 = \varepsilon \eta \frac{\Delta E (1-\mathcal{S})}{\tau_{\mathrm{n}}'}$ and $\varepsilon$ is the fraction of non-thermal phonons {\color{black} transmitted} from the carrier to the TES and $\eta$ is the fraction of non-thermal phonons {\color{black} transmitted} from the absorber to the carrier. $\Delta E$ is the energy deposited in the absorber and $\mathcal{S}$ is the fraction of energy converted into scintillation light and absorbed in the carrier. $\tau'_{\mathrm{n}}$ is the relaxation time constant of the non-thermal phonons, defined by
\begin{equation}
\label{eq:taun'}
\tau_{\mathrm{n}}' = \left( \frac{1}{\tau_{\mathrm{film}}} + \frac{1}{\tau_{\mathrm{c}}} + \frac{1}{\tau_{\mathrm{a}}}\right)^{-1}, 
\end{equation}

where $\tau_{\mathrm{film}}, \tau_{\mathrm{c}}, \tau_{\mathrm{a}}$ are the characteristic time {\color{black} constants} of non-thermal phonon thermalization in the TES, carrier and absorber, respectively. $P_\ell = \xi \frac{\Delta E \mathcal{S}}{\tau_\ell}$ is the power input to the TES due to absorption of scintillation light in the carrier, where $\xi$ is the fraction of non-thermal phonons generated in the carrier that reach the TES. The relaxation time constant of the non-thermal phonons generated by light absorption is
\begin{equation}
\tau_\ell = \left( \frac{1}{\tau_{\mathrm{film}}} + \frac{1}{\tau_{\mathrm{c}}}\right)^{-1}.
\end{equation}

The blue-highlighted carrier-TES subsystem in
Fig.~\ref{fig:det_scheme_carrier} illustrates the connection to the
original two-component model. The system {\color{black} reduces} to the two component original model (i.e. the system in Eq.~\eqref{eq:system} and its solution in Eq.~\eqref{eq:Franz}) if the absorber, thus also its couplings, and $P_\ell$, are removed from Eq.~\eqref{eq:extendedsystem}. {\color{black} This limit is obtained by setting $P_\ell=0$, $\eta=1$, and $\tau_{\mathrm{a}}\rightarrow\infty$. After this reduction, the remaining carrier component plays the role of the absorber in the original model, corresponding to the relabeling $\mathrm{c}\rightarrow\mathrm{a}$.} 

The differential system in Eq. \ref{eq:extendedsystem} has the structure,
\begin{equation}
\label{eq:diffeq}
	\begin{cases}
		\dot{\textbf{x}}(t) & = \textbf{A}\; \textbf{x} + \textbf{f}(t) \\
		\textbf{x}(t=0) & = \begin{pmatrix}
						T_{\mathrm{b}} & T_{\mathrm{b}} & T_{\mathrm{b}}
					     \end{pmatrix}^T
	\end{cases},
\end{equation}

where
\begin{equation}
\textbf{x}(t)  = 
	\begin{pmatrix}
	T_{\mathrm{e}}(t) & T_{\mathrm{c}} (t) & T_{\mathrm{a}} (t) 
	\end{pmatrix}^T.
\end{equation}

The system is a 3-dimensional non-homogeneous first order differential system, with constant coefficients. Its solution is

\begin{equation}
\textbf{x}(t) = e^{\textbf{A} \; t}\;  \textbf{c} + \int^{t}_{0} ds \; e^{\textbf{A}(t-s)} \; \textbf{f} (s), 
\end{equation}
where $\textbf{c}$ is a vector of coefficients fixed by initial conditions. To simplify the calculation of its analytical solution, we parameterize the system as follows. We label with $\lambda_1, \lambda_2$ and $\lambda_3$ the three eigenvalues of the ($3\times3$)-matrix $A$. The three eigenvectors $\textbf{v}_i$ can be found solving the eigenvalue system as function of $\lambda_i$,
\begin{equation}
\textbf{v}_i = (1, \alpha(\lambda_i), \beta(\lambda_i)) \equiv  (1, \alpha_i, \beta_i),
\end{equation}

where

\begin{equation}
	\begin{cases}
		\alpha_i & = \frac{C_{\mathrm{e}}}{G_{\mathrm{ec}}}\left(\frac{G_{\mathrm{ec}}}{C_{\mathrm{e}}}+\frac{G_{\mathrm{eb}}}{C_{\mathrm{e}}} +\lambda_i \right)\\
		\beta_i  & = - \alpha_i \frac{G_{\mathrm{ac}}}{C_{\mathrm{a}}}\left( -\frac{G_{\mathrm{ac}}}{C_{\mathrm{a}}} - \frac{G_{\mathrm{ab}}}{C_{\mathrm{a}}} -\lambda_i  \right)^{-1} 
	\end{cases}.
\end{equation}
It follows that the exponential matrix $e^{\textbf{A} \; t}$ is

\begin{equation}
\begin{aligned}
e^{\textbf{A} \; t} & = \begin{pmatrix}
				1 & 1 & 1\\
				\alpha_1 & \alpha _2 & \alpha_3\\
				\beta_1   & \beta _2    & \beta_3
            		     \end{pmatrix}
		     		\cdot e^{\textbf{J} \; t} \cdot
				\begin{pmatrix}
				1 & 1 & 1\\
				\alpha_1 & \alpha _2 & \alpha_3\\
				\beta_1   & \beta _2    & \beta_3
            		     \end{pmatrix}^{-1}  \\
		      & = \begin{pmatrix}
				1 & 1 & 1\\
				\alpha_1 & \alpha _2 & \alpha_3\\
				\beta_1   & \beta _2    & \beta_3
            		     \end{pmatrix}
		     		\cdot e^{\textbf{J} \; t} \cdot
				\begin{pmatrix}
				m_1 & \ell_1 & k_1\\
				m_2 & \ell_2 & k_2\\
				m_3 & \ell_3 & k_3\\
            		     \end{pmatrix}  \\
		      &	 = \begin{pmatrix} 
		                \sum_i e^{\lambda_i}m_i & \sum_i e^{\lambda_i}\ell_i & \sum_i e^{\lambda_i}k_i \\
				\sum_i \alpha_i e^{\lambda_i}m_i & \sum_i \alpha_i e^{\lambda_i}\ell_i & \sum_i \alpha_i e^{\lambda_i}k_i \\
				 \sum_i \beta_i e^{\lambda_i}m_i & \sum_i \beta_i e^{\lambda_i}\ell_i & \sum_i \beta_i e^{\lambda_i}k_i \\
            		     \end{pmatrix}
\end{aligned}
\end{equation}
with $\textbf{J} = \mathrm{diag}(\lambda_1,\lambda_2, \lambda_3)$ and $e^{\textbf{J} \; t} = \mathrm{diag}(e^{\lambda_1\; t} ,e^{ \lambda_2 \; t}, e^{\lambda_3\; t} )$. With this convenient parametrization the final solution for $T_{\mathrm{e}}$ is computed,
\begin{equation}
\label{eq:extendedmodel}
\Delta T_{\mathrm{e}}(t) = \sum_{i=1}^3 A_i \; [e^{\lambda_i\; t} - e^{-t/\tau_\ell}] + \sum_{i=1}^3 B_i \; [e^{\lambda_i\; t} - e^{-t/\tau'_{\mathrm{n}}}]
\end{equation}

with
\begin{equation}
\begin{cases}
\label{eq:amplitudes}
A_i & = \frac{ P_\ell}{\lambda_i + \frac{1}{\tau_\ell}}\left( \frac{m_i }{C_{\mathrm{e}}}  + \frac{\ell_i }{C_{\mathrm{c}}} \frac{1-\xi}{\xi} \right)\\   
 B_i  & = \frac{P_0}{\lambda_i + \frac{1}{\tau'_{\mathrm{n}}} } \left( \frac{m_i }{C_{\mathrm{e}}} +  \frac{\ell_i }{C_{\mathrm{c}}} \frac{1-\varepsilon}{\varepsilon}  + \frac{k_i }{C_{\mathrm{a}}} \frac{1-\eta}{\eta \varepsilon}  \right) 

\end{cases}.
\end{equation}

Four checkpoints are used through the solution of the system to compare it with the original model in~\cite{probst1995model}:
\begin{enumerate}
\item {\color{black} Imposing initial conditions, the coefficients are fixed by
$\mathbf{c}=\mathbf{x}(0)=(T_{\mathrm{b}},T_{\mathrm{b}},
T_{\mathrm{b}})^T$, as in the original model;}
\item {\color{black}The time-independent terms correspond to the bath temperature
$T_{\mathrm{b}}$, i.e., the equilibrium temperature in the absence of a
pulse. Therefore, the pulse signal is given by the TES temperature excursion
relative to the bath, $\Delta T_{\mathrm{e}}(t)\equiv T_{\mathrm{e}}(t)-T_{\mathrm{b}}=\Delta T_{\mathrm{eb}}(t)$ \footnote{This assumes that static heater
power and TES self-heating are neglected. }
;}
\item {\color{black} The final solution of the differential system can be written as the sum of differences of exponentials, because the amplitudes $A_i, B_i$ can be factorized, as in the original model};
\item The original model is recovered with $\eta = 1$, $\tau_{\mathrm{a}} \rightarrow \infty$ and $P_\ell = 0$. 
\end{enumerate}

From Eq.~\eqref{eq:extendedmodel} two non-thermal components, one for the A-part and one for the B-part, and four thermal components, two for the A-part and two for the B-part, can be identified.
 $\tau'_{\mathrm{n}}$, as well as $\tau_\ell$, can be considered as decay time constants of the non-thermal components and rise time constants of the thermal components. One of the three $\tau_i \equiv -(\lambda_i)^{-1}$ can be interpreted as the intrinsic response time of the TES, rise time constant of the non-thermal component, and the other two $\tau_i$ as the decay time constants of the two thermal components. If $\tau_1$ is assumed as the intrinsic response time of the TES, in analogy with the original model, Eq.~\eqref{eq:extendedmodel} can be written as

\begin{equation}
\label{eq:extendedmodel_phys}
\begin{aligned}
\Delta T_{\mathrm{e}}(t) ={}&
 A'_1 \left[e^{-t/\tau_{\ell}}
 -e^{-t/\tau_{\mathrm{in}}}\right]
+ A_2 \left[e^{-t/\tau_{\mathrm{t}_1}}
 -e^{-t/\tau_{\ell}}\right] \\
+& A_3 \left[e^{-t/\tau_{\mathrm{t}_2}}
 -e^{-t/\tau_{\ell}}\right]
+ B'_1 \left[e^{-t/\tau'_{\mathrm{n}}}
 -e^{-t/\tau_{\mathrm{in}}}\right] \\
+& B_2 \left[e^{-t/\tau_{\mathrm{t}_1}}
 -e^{-t/\tau'_{\mathrm{n}}}\right]
+ B_3 \left[e^{-t/\tau_{\mathrm{t}_2}}
 -e^{-t/\tau'_{\mathrm{n}}}\right].
\end{aligned}
\end{equation}

with
\begin{equation}
\begin{cases}
 A'_1  & = -A_1 \\
B'_1 & = - B_1
\end{cases},
\end{equation}

where $A_i$ and $B_i$ are defined in Eq.~\eqref{eq:amplitudes}, $\tau_{\mathrm{in}} = \tau_1$, $\tau_{\mathrm{t_1}} = \tau_2$ and $\tau_{\mathrm{t_2}} = \tau_3$.\\

\paragraph{A remark:}
The analytical solution for the amplitudes in Eq.~\eqref{eq:amplitudes} shows that the amplitudes $A_i$ are proportional to the power input, $P_\ell$, injected by non-thermal phonons produced in the carrier by the scintillation light absorption and transmitted to the TES, while the $B_i$ are proportional to the power input, $P_0$, injected by non-thermal phonons produced in the absorber and transmitted to the TES. It follows that Eq.~\eqref{eq:extendedmodel} is describing the TES temperature variation in the phonon channel, $\Delta T_{\mathrm{e}}$, induced by both the phonon and the light generated by a scattering in the absorber.  

\subsection{How does the model describe the observations?}
The solution in Eq.~\eqref{eq:extendedmodel_phys} contains additional thermal components which play the role of the empirical second thermal component which was introduced in~\cite{Angloher:2017sft}, hence it can explain the first observation related to the NaI pulse shape. Such second thermal component is predicted by the new model also if 100\% transmittance of the scintillation light from the carrier is assumed (that is $P_\ell = 0$ and $A_i = 0$).\\

{\color{black}
{\color{black}
The second observation was the absence of a nuclear recoil band in the $LY$ plot. 
The $LY$ used for particle discrimination is defined as
\begin{equation}
\label{eq:LY}
LY = \frac{E_{\ell}^{\mathrm{b}}}{E_{\mathrm{ph}}},
\end{equation}
where $E_{\ell}^{\mathrm{b}}$ is the energy reconstructed by the light detector, which in COSINUS is a silicon beaker equipped with a TES, and $E_{\mathrm{ph}}$ is the reconstructed phonon channel signal after a particle interaction.

In the composite detectors considered here, the phonon channel TES signal receives two contributions: one from phonons produced directly by the particle interaction in the absorber, $E_{\mathrm{p}}$, and one from non-thermal phonons generated in the carrier by absorbed scintillation light, $E_{\ell}^{\mathrm{c}}$. Thus,
\begin{equation}
\label{eq:Eph_composite}
E_{\mathrm{ph}} = E_{\mathrm{p}} + E_{\ell}^{\mathrm{c}}.
\end{equation}

The light energy measured by the external light detector, $E_{\ell}^{\mathrm{b}}$, is in general not identical to the light-induced contribution measured in the phonon channel, $E_{\ell}^{\mathrm{c}}$, since the beaker and carrier have different geometries and absorption efficiencies. However, to first approximation both quantities scale with the scintillation light produced in the NaI absorber, and we write
\begin{equation}
\label{eq:light_proportionality}
E_{\ell}^{\mathrm{b}} = \kappa E_{\ell}^{\mathrm{c}},
\end{equation}
with a constant proportionality factor $\kappa$. The measured $LY$ is then
\begin{equation}
\label{eq:LY_composite}
LY = \frac{\kappa E_{\ell}^{\mathrm{c}}}{E_{\mathrm{p}}+E_{\ell}^{\mathrm{c}}}.
\end{equation}

If the phonon channel amplitude is dominated by the light-induced contribution, $E_{\ell}^{\mathrm{c}}\gg E_{\mathrm{p}}$, this reduces to
\begin{equation}
\label{eq:LY2}
LY \simeq \kappa .
\end{equation}

In this regime, the $LY$ becomes approximately independent of the amount of scintillation light and thus of particle type. After the usual normalization of the $e^-/\gamma$ band to $LY=1$, neutron-induced events can populate the same band, explaining the absence of a distinct nuclear recoil band.

The light-induced contribution can dominate the phonon channel amplitude because the amplitudes of the pulse components are proportional to their corresponding power inputs. In particular, from Eq.~\eqref{eq:amplitudes}, the coefficients $A_i$ are proportional to $P_\ell$, while the coefficients $B_i$ are proportional to $P_0$. From Eq.~\eqref{eq:powerinputs},
\begin{equation}
P_\ell = \xi \frac{\Delta E\mathcal{S}}{\tau_\ell},
\qquad
P_0 = \varepsilon\eta \frac{\Delta E(1-\mathcal{S})}{\tau'_{\mathrm{n}}}
.
\end{equation}
Their ratio is
\begin{equation}
\frac{P_\ell}{P_0}
=
\frac{\xi}{\varepsilon\eta}
\frac{\mathcal{S}}{1-\mathcal{S}}
\frac{\tau'_{\mathrm{n}}}{\tau_\ell}.
\label{pl_over_po}
\end{equation}
Thus, even for $\mathcal{S}\ll1$, the light-induced component can dominate the phonon channel amplitude if its time scale is sufficiently shorter, $\tau_\ell \ll \tau'_{\mathrm{n}}$, and/or if the collection of light-induced non-thermal phonons into the TES is more favorable than the collection of absorber-generated non-thermal phonons into the TES, $\xi \gg \varepsilon\eta$. The absence of a separated nuclear recoil band already indicates that the detector operated in such a light-dominated phonon channel regime; this interpretation is quantitatively supported by the pulse-shape fits discussed in Sec.~\ref{sec:LEPD}.

}}

\section{A Light Enhanced Phonon Detector: Pulse Shape Studies} 
\label{sec:LEPD}
Eq.~\eqref{eq:extendedmodel} describes the TES response of a \textit{light-enhanced phonon detector} to phonons from the absorber and scintillation light absorbed in the carrier. We test this phenomenology by fitting the pulse shape model in Eq.~\eqref{eq:extendedmodel} to the pulses recorded using the composite detectors NaIce2 and NaIce6.

\begin{figure}[htbp]
\centering

\begin{subfigure}[t]{0.32\textwidth}
  \centering
  \includegraphics[
    width=\linewidth,
    height=0.72\linewidth
  ]{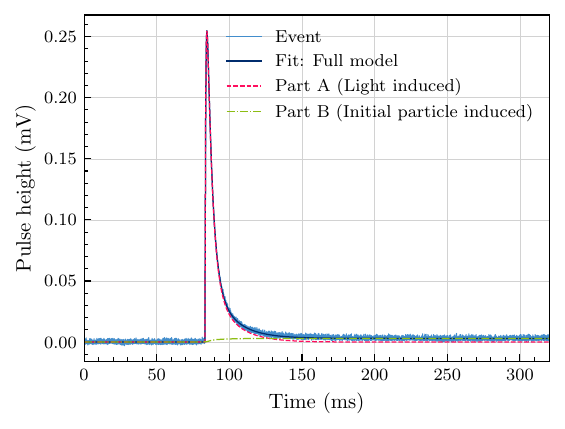}
  \caption{NaIce2}
  \label{fig:naice2}
\end{subfigure}
\hfill
\begin{subfigure}[t]{0.32\textwidth}
  \centering
  \includegraphics[
    width=\linewidth,
    height=0.72\linewidth
  ]{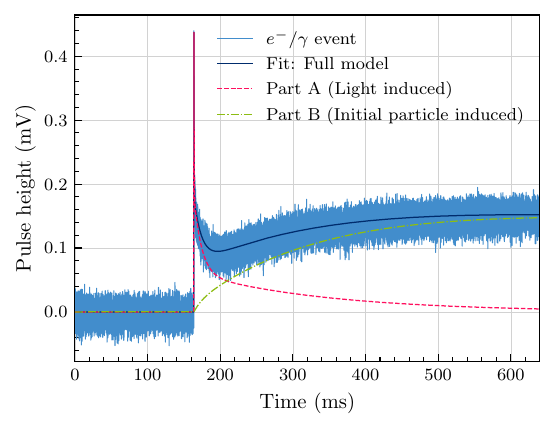}
  \caption{NaIce6 ($e^-/\gamma$)}
  \label{fig:naice6-egamma}
\end{subfigure}
\hfill
\begin{subfigure}[t]{0.32\textwidth}
  \centering
  \includegraphics[
    width=\linewidth,
    height=0.72\linewidth
  ]{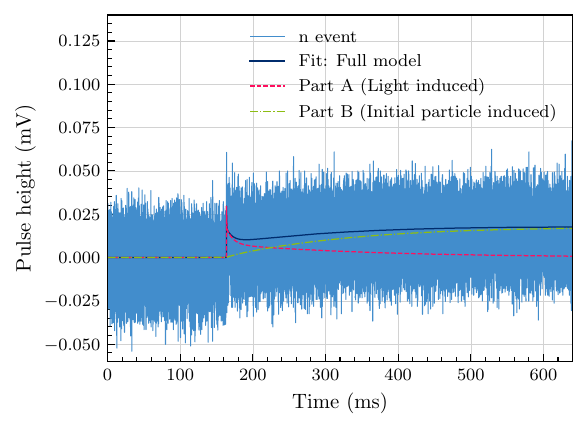}
  \caption{NaIce6 (neutron)}
  \label{fig:naice6-n}
\end{subfigure}

\caption{ Pulse shape fits for representative events in NaIce2 and NaIce6. For NaIce6, examples of an NTD-tagged $e^-/\gamma$ event and an NTD-tagged {\color{black}nuclear recoil} event are shown. The contributions of the two phonon populations (light-induced and {\color{black}initial particle}-induced) are indicated together with the total model prediction.}
\label{fig:three-panels}
\end{figure}

{\color{black}
Since the parameter space is high-dimensional and exhibits strong parameter degeneracies, we perform the fits within a Bayesian framework using BAT.jl~\cite{Schulz:2021BAT}. This approach enables sampling of the full posterior distribution, thereby accounting for parameter uncertainties and correlations.\footnote{{\color{black}The fitting procedure was developed in \cite{Master_Sarah}; a dedicated publication is in preparation.}} To improve convergence and reduce correlations among the fitted parameters, the model is fitted in a reparameterized form. Since the recorded waveforms are fitted directly in voltage units, the
temperature-like quantities in the model are expressed in
voltage-equivalent units. For the small pulse amplitudes considered here,
we assume a linear relation, to first order, between the TES temperature
variation and the readout voltage. No absolute voltage-to-temperature
calibration is applied. The free fit parameters are listed in
Tab.~\ref{FreeParams_3Comp}.

\begin{table}
\centering
\caption{Free parameters used in the fit. The source terms $a_i$ and
$b_i$ represent non-thermal power inputs divided by the heat capacity
of the driven component. They are therefore temperature rates in the
physical model and are reported here in voltage-equivalent units of
$\mathrm{mV}/\mathrm{ms}$.}
\renewcommand{\arraystretch}{1.1} 
\begin{tabular}{
>{\centering\arraybackslash}m{0.33\linewidth}
>{\centering\arraybackslash}m{0.52\linewidth}
>{\centering\arraybackslash}m{0.08\linewidth}
}
\hline
Free parameter & Description & Unit \\
\hline

$t_{\mathrm{onset}}$
& Pulse onset time
& $\mathrm{ms}$ \\

$\tau'_{\mathrm{n}}$ &
\multirow{2}{0.82\linewidth}{\centering Non-thermal relaxation times of absorber- and light-induced phonons}
& \multirow{2}{*}{$\mathrm{ms}$} \\
$\tau_{\ell}$ & & \\

$\frac{G_{\mathrm{ec}}}{C_{\mathrm{e}}}$ &
\multirow{5}{0.92\linewidth}{\centering Thermal response rates of the TES, carrier, and absorber for their pairwise couplings}
& \multirow{5}{*}{$\frac{1}{\mathrm{ms}}$} \\
$\frac{G_{\mathrm{ec}}}{C_{\mathrm{c}}}$ & & \\
$\frac{G_{\mathrm{ac}}}{C_{\mathrm{a}}}$ & & \\
$\frac{G_{\mathrm{ac}}}{C_{\mathrm{c}}}$ & & \\
$\frac{G_{\mathrm{eb}}}{C_{\mathrm{e}}}$ & & \\

$a_1=\frac{\xi\,\Delta E\,\mathcal{S}}{C_{\mathrm{e}}\,\tau_{\ell}}$ &
\multirow{2}{0.82\linewidth}{\centering Light-induced source terms in TES and carrier}
& \multirow{2}{*}{$\frac{\mathrm{mV}}{\mathrm{ms}}$} \\
$a_2=\frac{(1-\xi)\,\Delta E\,\mathcal{S}}{C_{\mathrm{c}}\,\tau_{\ell}}$ & & \\

$b_1=\frac{\varepsilon\,\eta\,\Delta E\,(1-\mathcal{S})}{C_{\mathrm{e}}\,\tau'_{\mathrm{n}}}$ &
\multirow{3}{0.62\linewidth}{\centering Absorber-induced source terms in TES, carrier, and absorber}
& \multirow{3}{*}{$\frac{\mathrm{mV}}{\mathrm{ms}}$} \\
$b_2=\frac{(1-\varepsilon)\,\eta\,\Delta E\,(1-\mathcal{S})}{C_{\mathrm{c}}\,\tau'_{\mathrm{n}}}$ & & \\
$b_3=\frac{(1-\eta)\,\Delta E\,(1-\mathcal{S})}{C_{\mathrm{a}}\,\tau'_{\mathrm{n}}}$ & & \\

\hline
\end{tabular}
\label{FreeParams_3Comp}
\end{table}

With this parameterization, we can still recover the non-thermal collection efficiencies $\eta$, $\varepsilon$, and $\xi$, the scintillation fraction $\mathcal{S}$, as well as the non-thermal relaxation times in the individual detector components. In particular, we infer $\tau_{\mathrm{a}}$, $\tau_{\mathrm{c}}$, and $\tau_{\mathrm{film}}$ for non-thermal phonons originating from an energy deposition in the absorber. For the non-thermal phonons generated by scintillation light absorption in the carrier, the corresponding carrier and TES-film relaxation times are denoted by $\tau_{\mathrm{c}}^{\ell}$ and $\tau_{\mathrm{film}}^{\ell}$, respectively. The uncertainties of these quantities, which are not fitted directly but derived from the sampled fit parameters, are obtained via Bayesian error propagation.

Example fits are shown in Fig.~\ref{fig:three-panels}. For NaIce2, a representative event is displayed. {\color{black}For NaIce6, neutron and
$e^-/\gamma$ events were identified from the light yield reconstructed
using the NTD phonon channel and the external light detector. Compared
with the TES, the NTD predominantly measures the thermal energy and
is less affected by the light-induced phonon component. Together
with its more linear response at higher energies, this provides a
reliable phonon energy estimator and enables separation of the two event
classes. One representative TES event from each class is shown.}

To enable a consistent comparison of the inferred parameters between the two event classes, we use identical prior distributions for neutron and $e^-/\gamma$ fits for all parameters that are not expected to depend on particle type (in particular the thermal response rates). Only the priors of the amplitude-related parameters $a_i$ and $b_i$
are chosen broader, since these parameters encode both the
event-to-event variation in deposited energy and its particle dependent
partition between scintillation light and initial phonons due to
quenching. We therefore adopt weakly informative priors to let the data constrain these contributions. Consistently, these parameters are the only ones that differ significantly between the two event classes, while the remaining parameters are compatible within uncertainties.} In all plots, the two phonon populations entering the model and the corresponding total model are shown.

A qualitative comparison shows that, in NaIce2, the absorber-phonon contribution is much less pronounced relative to the light-induced population than in NaIce6. This is in agreement with the expectation from the transmittance curves. The results of the pulse shape fits demonstrate that the approximation in Eq.~\eqref{eq:LY2} is justified, {\color{black} i.e. the phonon signal is dominated by scintillation light absorption in the carrier.}

The fit results most relevant for the following discussion are summarized in Tab.~\ref{fit_params}; the complete set of fit and derived parameters is reported in Tab.~\ref{tab:fit-params-all}.

\begin{table}[ht]
\caption{ Most relevant fit results for NaIce2, NaIce6
($e^-/\gamma$), and NaIce6 (neutron), reported as global posterior
modes with posterior standard deviations. Derived quantities are
obtained from the posterior samples using Bayesian error propagation.
Superscript $\ell$ denotes non-thermal relaxation times for phonons
generated by scintillation light absorption in the carrier.}

\centering

\begin{tabular}{c c c c}
\hline
Parameter & NaIce2 & NaIce6 ($e^-/\gamma$) & NaIce6 (neutron) \\
\hline
$\tau'_{\mathrm{n}}$ ($\mathrm{ms}$)     
& $(2.96 \pm 0.04)\times   10^{3}$
& $(2.96 \pm 0.06)\times   10^{3}$
& $(2.94 \pm 0.07)\times   10^{3}$ \\

$\tau_{\ell}$ ($\mathrm{ms}$)      
& $0.038 \pm 0.004$ 
& $0.023 \pm 0.003$ 
& $0.04 \pm 0.01$ \\

$\varepsilon$ 
& $0.016 \pm 0.003$ 
& $(3 \pm 2)\times  10^{-3}$ 
& $(5 \pm 2)\times  10^{-3}$ \\

$\eta$  
& $0.985 \pm 0.003$ 
& $0.86 \pm 0.04$ 
& $0.8 \pm 0.1$ \\

$\xi$  
& $0.015 \pm 0.002$ 
& $0.011 \pm 0.001$ 
& $(8 \pm 4)\times  10^{-3}$ \\

$\mathcal{S}$  
& $0.172 \pm 0.002$ 
& $0.023 \pm 0.002$ 
& $0.021 \pm 0.007$ \\

$\tau_{\mathrm{a}}$ ($\mathrm{ms}$) 
& $(2.0 \pm 0.5)\times  10^{5}$ 
& $(2.3 \pm 0.9)\times 10^{4}$ 
& $(3 \pm 4)\times 10^{4}$ \\

$\tau_{\mathrm{c}}$ ($\mathrm{ms}$) 
& $(3.04 \pm 0.04)\times 10^{3}$ 
& $(3.4 \pm 0.2)\times 10^{3}$ 
& $(3.6 \pm 0.5)\times 10^{3}$ \\

$\tau_{\mathrm{film}}$ ($\mathrm{ms}$) 
& $(2.0 \pm 0.4)\times 10^{5}$ 
& $(1.5 \pm 0.2)\times 10^{6}$ 
& $(9 \pm 5)\times 10^{5}$ \\

$\tau_{\mathrm{c}}^{\ell}$ ($\mathrm{ms}$) 
& $0.038 \pm 0.003$ 
& $0.024 \pm 0.003$ 
& $0.03 \pm 0.01$ \\

$\tau_{\mathrm{film}}^{\ell}$ ($\mathrm{ms}$) 
& $2.5 \pm 0.3$ 
& $2.1 \pm 0.3$ 
& $5 \pm 3$ \\

$\frac{A_{\max}}{B_{\max}}$  
& $83.7 \pm 0.9$ 
& $2.97 \pm 0.02$ 
& $1.7 \pm 0.3$ \\
\hline
\end{tabular}

\label{fit_params}
\end{table}

First, we note that the non-thermal decay time $\tau'_{\mathrm{n}}$ is very long for all detectors. This already indicates that non-thermal phonons do not thermalize in the TES on a short time scale. Inspecting the component-specific non-thermal time constants $\tau_{\mathrm{a}}$, $\tau_{\mathrm{c}}$, and $\tau_{\mathrm{film}}$, we find $\tau_{\mathrm{a}} > \tau_{\mathrm{c}}$ for all detectors. This implies that non-thermal phonons can thermalize in the carrier before they down-convert in the absorber, which is favorable for signal transmission across the absorber-carrier interface. In contrast, $\tau_{\mathrm{film}}$ is significantly larger than the corresponding carrier time constants. Hence, the collection and thermalization of non-thermal phonons in the TES film is inefficient. Consistently, the collection efficiency at the carrier-TES interface for absorber-generated non-thermal phonons, quantified by $\varepsilon$, is much smaller than the collection efficiency at the absorber-carrier interface, quantified by $\eta$.

For the non-thermal phonon population generated by absorbed scintillation light in the carrier, we find significantly different decay times. In particular, $\tau_{\mathrm{c}}^{\ell}$ and $\tau_{\mathrm{film}}^{\ell}$ are about five orders of magnitude shorter than the corresponding time constants for the phonon population originating from energy depositions in the absorber. This implies that non-thermal phonons produced in the carrier by light absorption are collected and thermalize much faster than those originating in the absorber. Interestingly, the corresponding collection efficiency at the carrier-TES interface, $\xi$, is of the same order as $\varepsilon$ for the absorber-induced phonon population, i.e., the interface efficiency itself appears comparable for both populations. 

The strong separation of time scales was not initially considered in the model and is a result of the fit. It provides a clear indication that the non-thermal phonon decay dynamics are strongly frequency dependent as theoretically expected \cite{MARSIGLIO2020168102}: the phonon spectrum produced by light absorption in the carrier differs from that produced by particle interactions in the absorber, resulting in different decay time constants. This is an important observation, as it suggests that the pulse shape fit is sensitive to non-thermal phonon frequencies, which could be used to gain further insight into phonon transport in cryogenic detectors and will be investigated in the future.

We finally discuss scintillation light related parameters, namely the scintillation fraction $\mathcal{S}$ (the fraction of the initial deposited energy that is converted into scintillation light and subsequently absorbed in the carrier) and the ratio $\frac{A_{\max}}{B_{\max}}$ between the peak amplitude of the light-induced contribution and that of the absorber-induced phonon contribution. The inferred $\mathcal{S}$ is nearly an order of magnitude smaller for NaIce6 than for NaIce2. This is consistent with the optical properties shown in Fig.~\ref{fig:transmission}: $\mathrm{Al_2O_3}$ (the NaIce6 carrier) exhibits substantially higher transmission in the relevant wavelength range than $\mathrm{CdWO_4}$ (the NaIce2 carrier), and therefore absorbs less scintillation light. The same trend is reflected in $\frac{A_{\max}}{B_{\max}}$, which is about a factor of $40$ larger for NaIce2 than for NaIce6.

This raises the question whether the fit results, and in particular the parameters $\mathcal{S}$ and $\frac{A_{\max}}{B_{\max}}$, can be used to discriminate neutron events from $e^-/\gamma$ events. Since the scintillation light yield for nuclear recoils is reduced with respect to electron recoils due to quenching, both quantities are, in principle, expected to provide particle-type information.

To test this hypothesis, we fit ten {\color{black}randomly selected} NaIce6 events tagged as $e^-/\gamma$ events and ten events tagged as neutrons by the NTD. {\color{black}We restricted the analysis to ten events per class because of the computational cost of the inference method and the convergence difficulties requiring manual tuning of BAT.jl parameters.}

{\color{black}
For each fit, we derive the quantity
$\mathcal{E}_{\mathrm{fit}}=
\frac{\Delta E}{C_{\mathrm{e}}}$ from the fitted
source-term amplitudes $a_i$ and $b_i$. Within the model,
$\mathcal{E}_{\mathrm{fit}}$ is a temperature-equivalent measure of
the total energy deposited by the particle in the absorber, referenced
to the TES heat capacity $C_{\mathrm{e}}$. In the physical model,
$\frac{\Delta E}{C_{\mathrm{e}}}$ has units of temperature. Since the
waveforms are fitted in voltage units, we report
$\mathcal{E}_{\mathrm{fit}}$ in mV as a voltage-equivalent energy
proxy. As the voltage-to-temperature conversion is detector dependent
and no absolute calibration is available, this quantity does not define
an absolute energy scale.}

\begin{figure}[htbp]
\centering

\begin{subfigure}[t]{0.49\textwidth}
  \centering
  \includegraphics[width=\linewidth]{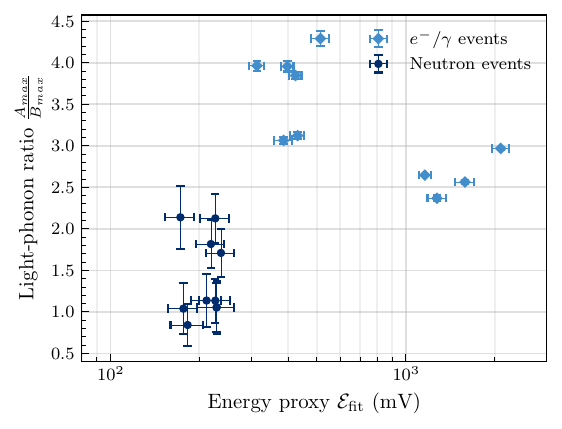}
  \caption{ NaIce6: $\frac{A_{\max}}{B_{\max}}$ vs\ the fit-derived
voltage-equivalent energy proxy $\mathcal{E}_{\mathrm{fit}}$. NTD-tagged $e^-/\gamma$ and {\color{black}nuclear recoil} events are shown in different colors. A separation based on $\frac{A_{\max}}{B_{\max}}$ appears feasible.}
  \label{fig:partdiscr-naice6-A}
\end{subfigure}\hfill
\begin{subfigure}[t]{0.49\textwidth}
  \centering
  \includegraphics[width=\linewidth]{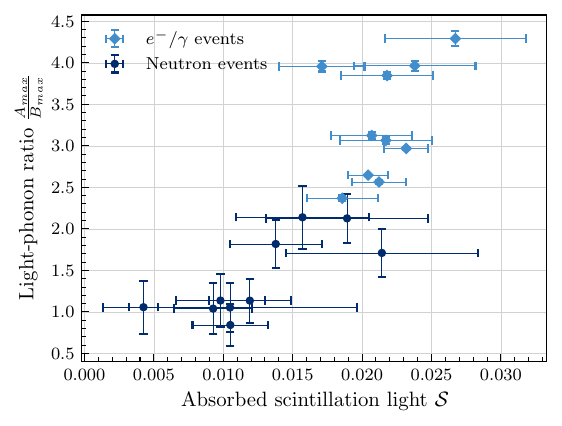}
  \caption{ NaIce6: $\frac{A_{\max}}{B_{\max}}$ vs\ $\mathcal{S}$. $\frac{A_{\max}}{B_{\max}}$ scales with $\mathcal{S}$, as expected; $\mathcal{S}$ shows no clear separation of event classes within uncertainties.}
  \label{fig:partdiscr-naice6-B}
\end{subfigure}

\vspace{3mm}

\begin{subfigure}[t]{0.49\textwidth}
  \centering
  \includegraphics[width=\linewidth]{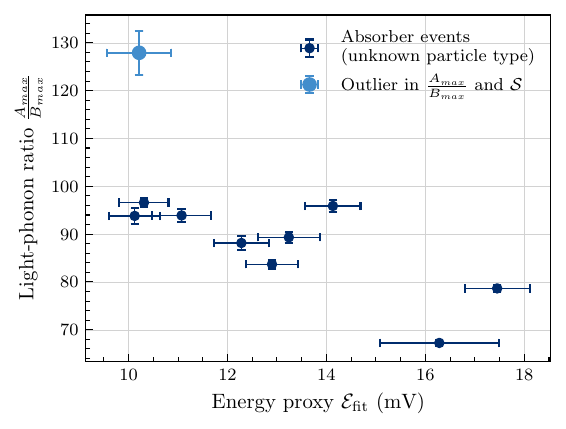}
  \caption{ NaIce2 (untagged): $\frac{A_{\max}}{B_{\max}}$ vs\ the fit-derived
voltage-equivalent energy proxy $\mathcal{E}_{\mathrm{fit}}$. One event appears as an outlier with larger $\frac{A_{\max}}{B_{\max}}$.}
  \label{fig:partdiscr-naice2-A}
\end{subfigure}\hfill
\begin{subfigure}[t]{0.49\textwidth}
  \centering
  \includegraphics[width=\linewidth]{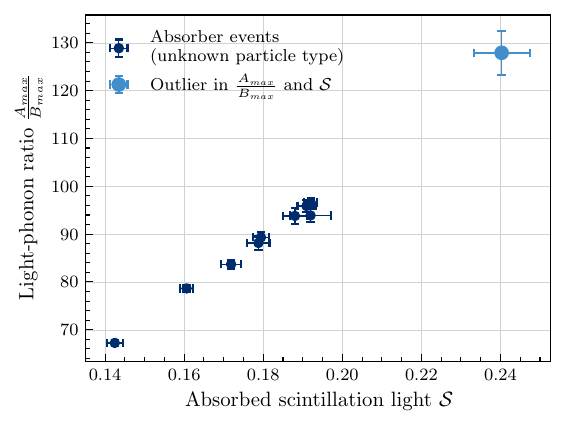}
  \caption{ NaIce2 (untagged): $\frac{A_{\max}}{B_{\max}}$ vs\ $\mathcal{S}$. The same outlier is visible particularly clearly.}
  \label{fig:partdiscr-naice2-B}
\end{subfigure}

\caption{ Particle discrimination observables derived from pulse shape fits for NaIce6 (top row; NTD-tagged $e^-/\gamma$ and {\color{black}nuclear recoil} events) and NaIce2 (bottom row; untagged events). Shown are the light-to-phonon ratio $\frac{A_{\max}}{B_{\max}}$ vs the fit-derived
voltage-equivalent energy proxy
$\mathcal{E}_{\mathrm{fit}}
=\frac{\Delta E}{C_{\mathrm{e}}}$ and vs the scintillation fraction $\mathcal{S}$ absorbed in the carrier. $\mathcal{E}_{\mathrm{fit}}$ is obtained from the fit and reported in
mV; it is not an absolutely calibrated deposited energy.}
\label{fig:partdiscr-all}
\end{figure}

We study $\frac{A_{\max}}{B_{\max}}$ as a function of
$\mathcal{E}_{\mathrm{fit}}$, and of the inferred scintillation fraction $\mathcal{S}$, to assess whether this single parameter coming directly from the fit alone provides discrimination power. The resulting scatter plots for NaIce6 are shown in Fig.~\ref{fig:partdiscr-all} (top row).

In Fig.~\ref{fig:partdiscr-naice6-A}, a clear particle-type dependence of $\frac{A_{\max}}{B_{\max}}$ is visible: neutron events exhibit smaller values, consistent with their reduced scintillation light output. Fig.~\ref{fig:partdiscr-naice6-B} shows that $\frac{A_{\max}}{B_{\max}}$ scales with $\mathcal{S}$. {\color{black}This is expected because the light-induced pulse amplitude scales with the power input proportional to $\mathcal{S}$, whereas the absorber-induced phonon amplitude scales with the power input proportional to $(1-\mathcal{S})$ (see Eq.~\eqref{pl_over_po}). Neglecting variations of the remaining response parameters, one therefore expects $\frac{A_{\max}}{B_{\max}} \propto \mathcal{S}/(1-\mathcal{S})$, which is approximately linear in the small $\mathcal{S}$ regime relevant here.}

However, $\mathcal{S}$ itself does not exhibit such a clear separation between neutron and $e^-/\gamma$ events in this data set. A plausible explanation is the comparatively large uncertainty of $\mathcal{S}$ as obtained from the fits. To assess whether $\mathcal{S}$ could provide particle-discrimination power in the case of smaller parameter uncertainties, and to test both observables on a data set without NTD tagging, we repeat the same analysis for 10 events recorded with NaIce2. The corresponding scatter plots are shown in Fig.~\ref{fig:partdiscr-all} (bottom row).

For NaIce2, particle discrimination is investigated on an untagged data set, i.e., no independent particle identification is available. Nevertheless, one event appears as a clear outlier in both $\frac{A_{\max}}{B_{\max}}$ and $\mathcal{S}$, while the remaining events cluster at significantly smaller values. This pattern is consistent with the outlier being an $e^-/\gamma$ event among only neutron-induced events, which would imply that $\mathcal{S}$ can provide discrimination power and could serve as an estimator that is more directly tied to the model than $\frac{A_{\max}}{B_{\max}}$.

To substantiate this interpretation, a comparison of neutron-calibration data with {\color{black}background} data and, in particular, fits to a substantially larger event sample would be required. At present, this is limited by the computational cost of Markov Chain Monte Carlo (MCMC)-based Bayesian inference {\color{black}and, more importantly, by the manual tuning of BAT.jl parameters required to achieve convergence.}

{\color{black}
We observe that the light-to-phonon ratio $\frac{A_{\max}}{B_{\max}}$ decreases with increasing $\mathcal{E}_{\mathrm{fit}}$ (Fig.~\ref{fig:partdiscr-naice2-A}). Within our pulse shape model this is accompanied by a decrease of the inferred scintillation fraction $\mathcal{S}$. 
Several effects could contribute to such a trend, including an energy-dependent scintillation non-proportionality of NaI (see, e.g., Ref.~\cite{Carlin:2025COSINE100}) and/or an energy-dependent light collection of the carrier. Since no absolute energy calibration is available for the present data set, a quantitative comparison to literature measurements is not possible. Clarifying the origin of this behaviour will require dedicated measurements with calibrated energies.}

\section{Conclusions and Outlook}
\label{sec:conclusions}

{\color{black}
We studied COSINUS composite NaI calorimeters and addressed two observations: pulse shapes cannot be described by the standard thermal model of Ref.~\cite{probst1995model}, and no nuclear recoil band is observed in the light yield plot. We showed that both can be explained by partial absorption of NaI scintillation light in the carrier crystal. The TES therefore measures a superposition of the absorber phonon signal and an additional light-induced phonon signal, turning the module into a \textit{light-enhanced phonon detector}.

We extended the model of Ref.~\cite{probst1995model} by including the carrier as a third thermal component and by adding a second non-thermal phonon population generated by light absorption in the carrier. The resulting analytical solution naturally contains an additional thermal component, explaining the empirical extra exponential term introduced in Ref.~\cite{Angloher:2017sft}. Bayesian fits to the data recorded with two detector modules validate the model and provide physically interpretable parameters, including phonon collection efficiencies and scintillation fractions.

Fit-derived observables were also investigated for particle discrimination. In NaIce6, NTD-tagged neutron and $e^-/\gamma$ events show separation in the light-to-phonon ratio $\frac{A_{\max}}{B_{\max}}$. In NaIce2, which had no NTD, an outlier appears consistently in both $\frac{A_{\max}}{B_{\max}}$ and $\mathcal{S}$. Together with the NTD-validated separation observed in NaIce6, this demonstrates that pulse shape fitting of the TES signal alone can provide particle discrimination.

Light-enhanced phonon detectors, combined with the pulse-shape analysis framework used here, could therefore provide a single-channel alternative to conventional two-channel cryogenic scintillating calorimeters. Future work will focus on quantifying the achievable discrimination power and energy threshold, adapting the model to other detector designs such as remoTES detectors, and accelerating the Bayesian analysis for application to larger data sets.}

\appendix

\section{Original pulse shape model}
We report the original pulse shape model by Pr\"obst et al.~\cite{probst1995model}. The pulse shape model presented in this work extends the original model to a composite design with three thermal detector components.

\begin{figure}[ht]
\begin{centering}
\includegraphics[width=0.5\linewidth]{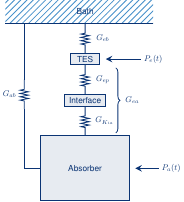}
\caption{Thermal model adapted from Fig.~6 of~\cite{probst1995model}.}
\label{fig:Franz_scheme}
\end{centering}
\end{figure}

\begin{equation}
\begin{cases}
\label{eq:system}
C_{\mathrm{e}} \frac{dT_{\mathrm{e}}}{dt} + G_{\mathrm{ea}}(T_{\mathrm{e}}-T_{\mathrm{a}})+G_{\mathrm{eb}}(T_{\mathrm{e}}-T_{\mathrm{b}}) & = P_{\mathrm{e}}(t)\\
C_{\mathrm{a}} \frac{dT_{\mathrm{a}}}{dt} + G_{\mathrm{ea}}(T_{\mathrm{a}}-T_{\mathrm{e}})+G_{\mathrm{ab}}(T_{\mathrm{a}}-T_{\mathrm{b}}) & = P_{\mathrm{a}}(t)\\
\end{cases}
\end{equation}

\begin{equation}
	\label{eq:power}
	\begin{cases}
		P_{\mathrm{e}}(t) & = \Theta (t) P_0 \; e^{-t/\tau_{\mathrm{n}}} \\
		P_{\mathrm{a}}(t) & = \frac{(1-\varepsilon)}{\varepsilon} P_{\mathrm{e}}(t)
	\end{cases}
\end{equation}

\begin{equation}
\label{eq:taun}
\tau_{\mathrm{n}} = \left( \frac{1}{\tau_{\mathrm{a}}} + \frac{1}{\tau_{\mathrm{film}}}\right)^{-1}
\end{equation}

\begin{equation}
\label{eq:diffeq}
	\begin{cases}
		\dot{\textbf{x}}(t) & = \textbf{A}\; \textbf{x} + \textbf{f}(t) \\
		\textbf{x}(t=0) & = \begin{pmatrix}
						T_{\mathrm{b}} &
						T_{\mathrm{b}}
					     \end{pmatrix}^T
	\end{cases}
\end{equation}

where,
\begin{equation}
\textbf{x}(t)  = 
	\begin{pmatrix}
	T_{\mathrm{e}}(t) & T_{\mathrm{a}} (t)
	\end{pmatrix}^T
\end{equation}

\begin{equation}
\label{eq:A}
\textbf{A}  = 
	\begin{pmatrix}
	-\frac{G_{\mathrm{eb}}+G_{\mathrm{ea}}}{C_{\mathrm{e}}} & \frac{G_{\mathrm{ea}}}{C_{\mathrm{e}}}\\
	 \frac{G_{\mathrm{ea}}}{C_{\mathrm{a}}} & -\frac{G_{\mathrm{ab}}+G_{\mathrm{ea}}}{C_{\mathrm{a}}}
	\end{pmatrix}
\end{equation}

\begin{equation}
	\textbf{f}(t) = 
		\begin{pmatrix} 
			\frac{ G_{\mathrm{eb}}T_{\mathrm{b}}}{C_{\mathrm{e}}}  + \frac{P_{\mathrm{e}}(t)}{C_{\mathrm{e}}} \\
		         \frac{ G_{\mathrm{ab}}T_{\mathrm{b}}}{C_{\mathrm{a}}} + \frac{P_{\mathrm{a}}(t)}{C_{\mathrm{a}}}
		\end{pmatrix} 
\end{equation}

\begin{equation}
\label{eq:Franz}
T_{\mathrm{e}}(t) = T_{\mathrm{b}} + \Theta(t) [A_{\mathrm{n}} \; (e^{-t/\tau_{\mathrm{n}}}-e^{ \lambda_1 \; t}) +A_{\mathrm{t}} \; (e^{\lambda_2 \; t}-e^{-t/\tau_{\mathrm{n}}})]
\end{equation}

\begin{align}
		A_{\mathrm{n}} & = - \frac{P_0 (s_{\mathrm{in}}-(G_{\mathrm{ab}}/C_{\mathrm{a}}))}{\varepsilon (s_{\mathrm{in}}-s_{\mathrm{t}} )(s_{\mathrm{in}}-s_{\mathrm{n}})}  \left( \frac{s_{\mathrm{t}} - (G_{\mathrm{ab}}/C_{\mathrm{a}})}{G_{\mathrm{eb}}-(C_{\mathrm{e}}/C_{\mathrm{a}})G_{\mathrm{ab}}}  - \frac{\varepsilon}{C_{\mathrm{e}}} \right)\\
		A_{\mathrm{t}} & = \frac{P_0 (s_{\mathrm{t}}-(G_{\mathrm{ab}}/C_{\mathrm{a}}))}{\varepsilon (s_{\mathrm{t}}-s_{\mathrm{in}} )(s_{\mathrm{t}}-s_{\mathrm{n}})}  \left( \frac{s_{\mathrm{in}} - (G_{\mathrm{ab}}/C_{\mathrm{a}})}{G_{\mathrm{eb}}-(C_{\mathrm{e}}/C_{\mathrm{a}})G_{\mathrm{ab}}}  - \frac{\varepsilon}{C_{\mathrm{e}}} \right)
\end{align}
with $s_i=1/\tau_i$ and $\lambda_i=-1/\tau_i$.

\section{Fit Parameters}
The complete set of fit and derived parameters for NaIce2 and
NaIce6  is reported in Tab.~\ref{tab:fit-params-all}.

\begin{table}[htb!]
\caption{Complete set of fit and derived parameters for NaIce2 and
NaIce6, given as global posterior modes with posterior standard
deviations. Temperature-like quantities are reported in
voltage-equivalent units because the waveforms were fitted directly
in voltage units.}
\label{tab:fit-params-all}

\centering

\begin{tabular}{c c c c}
\hline
Parameter & NaIce2 & NaIce6 ($e^-/\gamma$) & NaIce6 (neutron) \\
\hline

$t_{\mathrm{onset}}$ ($\mathrm{ms}$)
& $83.184 \pm 0.004$
& $163.790 \pm 0.002$
& $163.67 \pm 0.03$ \\

$\tau'_{\mathrm{n}}$ ($\mathrm{ms}$)
& $(2.96 \pm 0.04)\times  10^{3}$
& $(2.96 \pm 0.06)\times  10^{3}$
& $(2.94 \pm 0.07)\times  10^{3}$ \\

$\tau_{\ell}$ ($\mathrm{ms}$)
& $0.038 \pm 0.004$
& $0.023 \pm 0.003$
& $0.04 \pm 0.01$ \\

$a_1$ ($\frac{\mathrm{mV}}{\mathrm{ms}}$)
& $0.92 \pm 0.07$
& $23 \pm 2$
& $1.0 \pm 0.5$ \\

$a_2$ ($\frac{\mathrm{mV}}{\mathrm{ms}}$)
& $39 \pm 4$
& $200 \pm 40$
& $22 \pm 4$ \\

$b_1$ ($\frac{\mathrm{mV}}{\mathrm{ms}}$)
& $(5.9 \pm 0.9)\times  10^{-5}$
& $(2 \pm 1)\times  10^{-3}$
& $(3 \pm 1)\times  10^{-4}$ \\

$b_2$ ($\frac{\mathrm{mV}}{\mathrm{ms}}$)
& $(2.4 \pm 0.2)\times  10^{-3}$
& $0.05 \pm 0.01$
& $(8 \pm 6)\times  10^{-3}$ \\

$b_3$ ($\frac{\mathrm{mV}}{\mathrm{ms}}$)
& $(7 \pm 1)\times  10^{-5}$
& $(8 \pm 2)\times  10^{-3}$
& $(2 \pm 1)\times  10^{-3}$ \\

$\frac{G_{\mathrm{ec}}}{C_{\mathrm{e}}}$ ($\frac{1}{\mathrm{ms}}$)
& $0.38 \pm 0.03$
& $0.16 \pm 0.02$
& $0.09 \pm 0.04$ \\

$\frac{G_{\mathrm{ec}}}{C_{\mathrm{c}}}$ ($\frac{1}{\mathrm{ms}}$)
& $0.26 \pm 0.01$
& $0.015 \pm 0.001$
& $0.010 \pm 0.007$ \\

$\frac{G_{\mathrm{ac}}}{C_{\mathrm{a}}}$ ($\frac{1}{\mathrm{ms}}$)
& $0.104 \pm 0.003$
& $0.036 \pm 0.003$
& $0.05 \pm 0.01$ \\

$\frac{G_{\mathrm{ac}}}{C_{\mathrm{c}}}$ ($\frac{1}{\mathrm{ms}}$)
& $0.067 \pm 0.001$
& $0.051 \pm 0.002$
& $0.051 \pm 0.006$ \\

$\frac{G_{\mathrm{eb}}}{C_{\mathrm{e}}}$ ($\frac{1}{\mathrm{ms}}$)
& $1.16 \pm 0.05$
& $3.9 \pm 0.2$
& $4.0 \pm 0.4$ \\

$\varepsilon$
& $0.016 \pm 0.003$
& $(3 \pm 2)\times  10^{-3}$
& $(5 \pm 2)\times  10^{-3}$ \\

$\eta$
& $0.985 \pm 0.003$
& $0.86 \pm 0.04$
& $0.8 \pm 0.1$ \\

$\xi$
& $0.015 \pm 0.002$
& $0.011 \pm 0.001$
& $(8 \pm 4)\times  10^{-3}$ \\

$\mathcal{S}$
& $0.172 \pm 0.002$
& $0.023 \pm 0.002$
& $0.021 \pm 0.007$ \\

$\tau_{\mathrm{a}}$ ($\mathrm{ms}$)
& $(2.0 \pm 0.5)\times  10^{5}$
& $(2.3 \pm 0.9)\times  10^{4}$
& $(3 \pm 4)\times  10^{4}$ \\

$\tau_{\mathrm{c}}$ ($\mathrm{ms}$)
& $(3.04 \pm 0.04)\times  10^{3}$
& $(3.4 \pm 0.2)\times  10^{3}$
& $(3.6 \pm 0.5)\times  10^{3}$ \\

$\tau_{\mathrm{film}}$ ($\mathrm{ms}$)
& $(2.0 \pm 0.4)\times  10^{5}$
& $(1.5 \pm 0.2)\times  10^{6}$
& $(9 \pm 5)\times  10^{5}$ \\

$\tau_{\mathrm{c}}^{\ell}$ ($\mathrm{ms}$)
& $0.038 \pm 0.003$
& $0.024 \pm 0.003$
& $0.03 \pm 0.01$ \\

$\tau_{\mathrm{film}}^{\ell}$ ($\mathrm{ms}$)
& $2.5 \pm 0.3$
& $2.1 \pm 0.3$
& $5 \pm 3$ \\

$\mathcal{E}_{\mathrm{fit}}
=\frac{\Delta E}{C_{\mathrm{e}}}$
($\mathrm{mV}$)
& $12.9 \pm 0.5$
& $(2.1 \pm 0.1)\times  10^{3}$
& $240 \pm 30$ \\

$\frac{A_{\max}}{B_{\max}}$
& $83.7 \pm 0.9$
& $2.97 \pm 0.02$
& $1.7 \pm 0.3$ \\

\hline
\end{tabular}%

\end{table}

\section*{Acknowledgments}
The COSINUS collaboration gratefully acknowledges the directors and staff of the National Laboratory of Gran Sasso, as well as the technical staff of our home institutions, for their continuous support.
This work was supported by the Max-Planck-Institut für Physik, the Marietta Blau Institute for Particle Physics of the Austrian Academy of Sciences, the TU Wien, the National Institute of Nuclear Physics, and the Helsinki Institute of Physics. We acknowledge financial support from the Klaus Tschira Foundation, {\color{black} the Austrian Science Fund FWF through the Sonderforschungsbereiche (SFB) (grant DOI: 10.55776/I6956), the FWF through the Scies4Free programme (grant DOI: 10.55776/DFH66)}, the FWF (grant DOI: 10.55776/PAT1239524), the Research Council of Finland (grant no. 371542), the Finnish Graduate School in Particle and Nuclear Physics (Doctoral Education Pilot) and the Vilho, Yrjö and Kalle Väisälä Fund.

\bibliographystyle{JHEP}
\bibliography{biblio}
\end{document}